\documentstyle[aps,prb,epsf]{revtex} 

\newcommand{\subscrpt}[2]{{#1}_{{#2}}}

\newcommand{\bfc}{{\bf c}}
\newcommand{\bfcsub}[1]{\subscrpt{\bfc}{#1}}
\newcommand{\bfci}{\bfcsub{i}}

\newcommand{\bfx}{{\bf x}}

\newcommand{\bfxt}{(\bfx, t)}

\def\bfsigma{\mbox{\boldmath $\sigma$}}

\newcommand{\bge}{\begin{equation}}
\newcommand{\ee}{\end{equation}}
\newcommand{\bgc}{\begin{center}}
\newcommand{\ec}{\end{center}}
\newcommand{\bgea}{\begin{eqnarray}}
\newcommand{\eea}{\end{eqnarray}}
\newcommand{\bgeas}{\begin{eqnarray*}}
\newcommand{\eeas}{\end{eqnarray*}}

\newcommand{\hcc}{H_{\mbox{\scriptsize cc}}}
\newcommand{\hcd}{H_{\mbox{\scriptsize cd}}}
\newcommand{\hdc}{H_{\mbox{\scriptsize dc}}}
\newcommand{\hdd}{H_{\mbox{\scriptsize dd}}}

\begin{document}

\title{
\begin{flushleft}
{\footnotesize OUTP-97185}\\
{\footnotesize BU-CCS-970501}\\[0.3in]
\end{flushleft}
{\bf Lattice-Gas Simulations of Minority-Phase Domain Growth in
Binary Immiscible and Ternary Amphiphilic Fluids}
}
\author{
Florian W. J. Weig\footnote{Present address:
Ludwig-Maximilians-Universit\"{a}t, Theoretische Physik,
Theresienstra\ss e 37, D-80333 M\"{u}nchen, Germany, 
\mbox{florian.weig@extern.lrz-muenchen.de} }\\
{\small \sl Department of Theoretical Physics,}\\
{\small \sl Oxford University,}\\
{\small \sl 1 Keble Road, Oxford OX1 3NP, U.K.}\\
{\small \tt weig@thphys.ox.ac.uk}\\
Peter V. Coveney\footnote{Author to whom correspondence should be addressed}\\
{\small \sl Schlumberger Cambridge Research,}\\
{\small \sl High Cross, Madingley Road, Cambridge CB3 0EL, U.K.}\\
{\small \sl and}
{\small \sl Department of Theoretical Physics,}\\
{\small \sl Oxford University,}\\
{\small \sl 1 Keble Road, Oxford OX1 3NP, U.K.}\\
{\small \tt coveney@cambridge.scr.slb.com}\\
Bruce M. Boghosian\\
{\small \sl Center for Computational Science, Boston University}\\
{\small \sl 3 Cummington Street, Boston, Massachusetts 02215, U.S.A.}\\
{\small \tt bruceb@bu.edu}\\
[0.3cm]
}
\date{\today}
\maketitle

\begin{abstract}
  We investigate the growth kinetics of binary immiscible fluids and 
  emulsions in two dimensions using a hydrodynamic
  lattice-gas model. We perform off-critical quenches in the
  binary fluid case and find that the domain size within the minority 
  phase grows algebraically with time in accordance with theoretical 
  predictions. In the late time regime we find a growth exponent
  $n = 0.45 $ over a wide range of concentrations, in good agreement
  with other simulations. In the early time regime we find
  no universal growth exponent but a strong dependence on the
  concentration of the minority phase. In the ternary
  amphiphilic fluid case the kinetics of self assembly of the droplet phase
  are studied for the first time. At low surfactant concentrations, we
  find that, after an early algebraic growth, a nucleation regime
  dominates the late-time kinetics, which is enhanced by an
  increasing concentration of surfactant. 
  With a further increase in the concentration of surfactant, we see a
  crossover to logarithmically slow growth, and finally saturation of
  the oil droplets, which we fit phenomenologically to a stretched
  exponential function. Finally, the transition between the droplet and 
  the sponge phase is studied.  

  \noindent PACS numbers: 82.70.-y;05.70.Lm; 47.20.Hw; 64.60.Qb

\end{abstract}

\section{Introduction}

The fascinating effects caused by the introduction of
amphiphilic molecules into a system of oil and water have been subject
to considerable study for many years. For a general review of the
complex structures that arise due to the particular physical and
chemical properties of surfactant molecules, see Gelbart {\em et
al.}~\cite{bib:grb} or Gompper and Schick~\cite{bib:gs}. One major
feature of these systems is that the usual oil-water interfacial
tension is dramatically lowered by the presence of surfactant, a
phenomenon that underpins much of the interest in such self-assembling
amphiphilic structures. Making use of a hydrodynamic
lattice-gas model~\cite{bib:bce}, we demonstrate in the present paper that
we are able to consistently simulate growth kinetics in both
immiscible fluids and emulsions.

The growth kinetics of binary immiscible fluids quenched into the 
coexistence region have been studied extensively in the last decade. 
It is widely accepted that phase separation in these systems can be described 
in terms of a single quantity, the time dependent average domain size $R(t)$. 
In the regime of sharp domain walls, it is known that this quantity follows
algebraic growth laws, that is $R(t) \sim t^{n}$~\cite{bib:f,bib:bray}.
Much effort has been put into the determination of the growth exponent $n$ in
the case of deep critical quenches. Recent results
~\cite{bib:veto,bib:lwac,bib:cn,bib:kurao,bib:ecb} on two-dimensional fluids
confirm the presence of two regimes, the 
viscous regime~\cite{bib:smgg} with $n= \frac{1}{2}$ and the hydrodynamic 
or inertial regime~\cite{bib:f2} with $n= \frac{2}{3}$, which
dominate the early and late time behaviour respectively.   
Reports of an early time regime~\cite{bib:oosyb}, obeying the
Lifshitz-Slyozov evaporation-condensation mechanism with $n =
\frac{1}{3}$ are believed to be due to the absence of fluctuations in
the lattice-Boltzmann models employed.

In the case of off-critical quenches, however, comparatively few studies have 
been made and the situation is far from clear. Theoretical approaches
excluding hydrodynamics and fluctuations by Yao {\em et
al.}~\cite{bib:yegg} found
modified Lifshitz-Slyozov growth with $R(t)^3 = A + Bt$, where $A$ and
$B$ depend on the concentration of the minority phase. This leads to
a time-dependent growth exponent, which only asymptotically reaches $n
= \frac{1}{3}$. Numerical simulations by Chakrabarti
{\em et al.}~\cite{bib:chtg} confirmed this picture.
Implicitly including hydrodynamics, Furukawa~\cite{bib:f2} calculated
a value of the growth exponent of $n = \frac{2}{d+2}$ due to inertial
friction, which degenerates for $d = 2$ with the dissipative friction
regime ($n = \frac{1}{d}$) to $n = \frac{1}{2}$.
~\footnote{There has been some confusion about this value in
the literature, but it is only in $d = 3$ that Furukawa predicts $n =
\frac{2}{5}$, compare his review article~\cite{bib:f}.}
 San Miguel {\em et al.}~\cite{bib:smgg} predicted a droplet
coalescence growth with $n = \frac{1}{2}$, followed by a
Lifshitz-Slyozov evaporation-condensation mechanism with $n =
\frac{1}{3}$. But since this mechanism is suppressed by the
hydrodynamic modes in critical quenches, we believe that a
Lifshitz-Slyozov growth can only occur, if at all, before the regime with
$n = \frac{1}{2}$ in off-critical quenches as well.
Although some numerical simulations were undertaken so far using a
variety of techniques, none of them reported a growth exponent of $n =
\frac{1}{2}$ or $n = \frac{1}{3}$. Lattice-Boltzmann simulations by
Chen and Lookman~\cite{bib:cl} gave $n = 0.40 \pm 0.03 $, while Osborn
{\em et al.}~\cite{bib:oosyb} found $n=0.28 \pm 0.02$ in the high
viscosity and $n = 0.56 \pm 0.03$ in the low viscosity or hydrodynamic
regime. Langevin modelling by Wu {\em et al.}~\cite{bib:walc} gave
$n = 0.46 \pm 0.02$. Coveney and Novik~\cite{bib:cn} studied
quenches using dissipative particle dynamics and reported
$n = 0.47 \pm 0.02$. All of these simulations dealt with only one or
two values of minority phase concentration. Recently, Velasco and
Toxvaerd~\cite{bib:veto} using molecular dynamics methods measured the
domain growth for three different minority phase concentrations and
found that the growth exponent dropped continuously from $n \approx
0.45$ over $n \approx 0.33$ to a very slow growth (a crude measurement
in their data, Fig. 4 of~\cite{bib:veto} gives $n \approx 0.15 - 0.2$).
The latter was
explained by the authors to be due to the lack of statistical
significance. An earlier study by the same authors~\cite{bib:veto2}
had found a growth exponent of $n = 0.37$. Latest experiments on thin
films of off-critical polymer solutions by Haas and
Torkelson~\cite{bib:hato} found ``best-fit'' growth exponents ranging
from $n = 0.34$ up to $n = 0.42$, which were interpreted as a
Lifshitz-Slyozov growth with $n = 0.33$. Summing up all the results
obtained so far, we can only conclude that no consistent picture of
the growth laws in off-critical quenches has been established yet.
The strong variance of the reported growth exponents might indicate
that several rival scaling regimes lead to the measuring of an
exponents that are still in crossover regimes.
Recently Corberi {\em et al.}~\cite{bib:ccz} speculated about a
possible candidate for a very early time regime. Caused by the competition of
various fixed points, this regime could be visible in off-critical quenches
and would lead to a growth exponent of $n = \frac{1}{4}$.

In the present paper we study extensively the dependence of $n$ on
the concentration of the minority phase at different
temperatures. Starting from known behaviour in the critical quench 
we find strong evidence for a late time growth regime with $n = 0.45$,
which is stable over a wide range of concentrations. For early times
we do not find such a distinct regime, but a continuously
varying growth exponent. Possible reasons for this result are discussed.  

The kinetics of domain growth of ternary amphiphilic fluids, where
the presence of surfactant causes a drop in the interfacial tension
between the two otherwise immiscible fluids, for example oil and
water, is a comparatively new field of study. In the case of
microemulsions, for which there is a sufficient quantity of surfactant
present, the domains in these amphiphilic systems exhibit a preferred
length scale, and hence scale invariance must break down.
As a consequence of this we no longer expect algebraic
power growth laws in the late time kinetics. Instead, the requirement
that surfactant molecules sit at oil-water interfaces will lead to a
saturation of domain growth. Previous work studied the domain growth
using numerical integration of Landau-Ginzburg models, for example the
hybrid model of Kawakatsu {\em et al.}~\cite{bib:kkfok} and the 
two-local-order parameter model of Laradji and 
coworkers~\cite{bib:lggz,bib:lhgz}.
These models do not include hydrodynamic effects and find that
surfactant modifies the kinetics from the binary $n = \frac{1}{3}$
algebraic exponent to a slow growth that may be logarithmic in time. 
More recently Laradji {\em et al.}~\cite{bib:lmtz} have modelled phase
separation in the presence of surfactant using a very simple
molecular dynamics model, which implicitly includes hydrodynamic
forces. These authors found that such systems exhibit nonalgebraic, 
slow growth dynamics and a crossover scaling form, which describes
the change from the domain growth in pure binary immiscible fluids to
slower growth which occurs when surfactant is present. 
Emerton {\em et al.}~\cite{bib:ecb}, using a lattice-gas automaton
model~\cite{bib:bce} which implicitly includes fluctuations and
hydrodynamics, found with increasing surfactant concentration a
crossover from algebraic growth with $n = \frac{2}{3}$ to
$n = \frac{1}{2}$, then to logarithmic growth and finally to
saturation, which they fitted phenomenologically to a
stretched exponential function. 

All the papers mentioned above studied the kinetics of domain growth
within the sponge phase, the two dimensional equivalent to the
bicontinuous phase, where equal amounts of oil and water are present.
No work has been undertaken so far to study self-assembly kinetics in
the ``off-critical'' or droplet microemulsion phase. However, with our
amphiphilic lattice-gas model, we can readily access all the different
microemulsion phases; indeed, it has been shown that this model is
able to successfully simulate the droplet phase~\cite{bib:bce}.
 
The purpose of the present paper is therefore to make a detailed,
quantitative study of domain growth in off-critical quenches within
binary immiscible and ternary amphiphilic fluids.
Our paper is organized as follows: Sec.~\ref{sec:lgam} briefly reviews
the lattice-gas model we are using. In Sec.~\ref{sec:bif} we report our
results of off-critical quenches in binary fluids. The results of our
simulations in the droplet phase are given in Sec.~\ref{sec:tern}.
In Sec.~\ref{sec:trans} we look at the transition between droplet and
sponge phases. Finally, in Sec.~\ref{sec:dac}, we discuss our results.

\section{The Lattice-Gas Automaton Model}
\label{sec:lgam}

Our lattice-gas model is based on a microscopic particulate format that
allows us to include dipolar surfactant molecules alongside the basic
oil and water particles~\cite{bib:bce}. In this paper we are concerned
only with a two-dimensional version of the model, though an extension to
$3D$ is currently underway~\cite{bib:toappear}. Working on a triangular
lattice with lattice vectors $\bfci$ ($i=1,\ldots,6$) and periodic
boundary conditions, the state of the
$2D$ model at site $\bfx$ and time $t$ is completely specified by the
occupation numbers $n_i^{\alpha}(\bfx,t)\in\{0,1\}$ for particles of
species $\alpha$ and velocity $(\bfci/ \Delta t)$.

The evolution of the lattice gas for one timestep takes place in two
substeps.  In the {\it propagation} substep the particles simply move
along their corresponding lattice vectors.  In the {\it collision}
substep the newly arrived particles change their state in a manner that
conserves the mass of each species as well as the total $D$-dimensional
momentum.

We allow for two immiscible species which, following convention, we
often represent by colours: $\alpha=B$ (blue) for water, and $\alpha=R$
(red) for oil, and we define the {\it colour charge} of a particle moving
in direction $i$ at position $\bfx$ at time $t$ as $q_i\bfxt\equiv
n_i^R\bfxt-n_i^B\bfxt$.  Interaction energies between outgoing particles
and the total colour charge at neighbouring sites can then be calculated
by assuming that a colour charge induces a {\it colour potential} $\phi
(r)= q f(r)$, at a distance $r$ away from it, where $f(r)$ is some
function defining the type and strength of the potential.

To extend this model to amphiphilic systems, we also introduce a third
(surfactant) species $S$, and the associated occupation number
$n_i^S\bfxt$, to represent the presence or absence of a surfactant
particle.  Pursuing the electrostatic analogy, the surfactant particles,
which generally consist of a hydrophilic portion attached to a
hydrophobic (hydrocarbon) portion, are modelled as {\it colour dipole
  vectors}, $\bfsigma_i\bfxt$. As a result, the three-component model
includes three additional interaction terms, namely the colour-dipolar,
the dipole-colour and the dipole-dipole interactions.

Note that in order to incorporate the most general form of interaction
energy within our model system, we introduce a set of coupling constants
$\alpha, \mu, \epsilon, \zeta$, in terms of which the total interaction
energy can be written as
\bge
\Delta H_{\mbox{\scriptsize int}}
       =   \alpha \Delta \hcc +
         \mu \Delta \hcd +
         \epsilon \Delta \hdc +
         \zeta \Delta \hdd.
\label{eq:tiw}
\ee
These terms correspond, respectively, to the relative immiscibility of oil
and water, the tendency of surrounding dipoles to bend round oil or
water particles and clusters, the propensity of surfactant molecules to
align across oil-water interfaces and a contribution from pairwise
(alignment) interactions between surfactant.  In the present paper we
analyze domain growth of off-critical quenches within both binary and
ternary systems and consequently the coefficients with which we are
most concerned are $\alpha$ for the binary fluid case and $\epsilon$
and $\mu$ for the ternary fluid case.

The collision process of the algorithm consists of enumerating the
outgoing states allowed by the conservation laws, calculating the total
interaction energy for each of these, and then, following the ideas of
Chan and Liang~\cite{bib:cal} (see also Chen {\em et al.}~\cite{bib:chen}),
forming Boltzmann weights
\bge
e^{-\beta \Delta H},
\label{eq:bd}
\ee
where $\beta$ is an inverse temperature-like parameter.  The
post-collisional outgoing state and dipolar orientations can then be
obtained by sampling from the probability distribution formed from these
Boltzmann weights; consequently the update is a stochastic Monte-Carlo
process. The dipolar orientation streams with surfactant particles in
the usual way.

The parameter space of our model has certain important pairwise limits. With
no surfactant in the system, Eq.~(\ref{eq:tiw}) reduces to the
colour-colour interaction term only, which we note to be exactly
identical to the expression for the total
colour work used by Rothman and Keller~\cite{bib:rk} to model immiscible
fluids.  Correspondingly, with no oil in the system we are free to
investigate the formation and dynamics of the structures that are known
to form in binary water-surfactant solutions.  Indeed, in the original
paper Boghosian {\em et al.}~\cite{bib:bce} investigated both of 
these limits.  In the limit of no surfactant they obtained immiscible fluid 
behaviour similar to that observed by Rothman and Keller, and for the
case of no oil in the system they found evidence for the existence of
micelles and for a critical micelle concentration.  
It could be shown that this model exhibits the correct $2D$
equilibrium microemulsion phenomenology for both binary and ternary
phase systems using a combination of visual and
analytic techniques; various experimentally observed self-assembling
structures, such as the droplet and sponge microemulsion phases,
form in a consistent manner as a result of adjusting the relative
amounts of oil, water and amphiphile in the system. The presence of
sufficient surfactant in the system is shown to halt the expected phase
separation of oil and water, and this is achieved without altering the
coupling constants from values that produce immiscible behaviour in the
case of no surfactant.
Later studies~\cite{bib:ecb} showed that the model exhibits the correct
dynamic behaviour in the case of critical quenches of binary fluids and 
was able to give quantitative results for the self-assembling
kinetics in the microemulsion phase of a ternary oil-water-surfactant fluid.

\section{Phase separation in binary immiscible fluids}
\label{sec:bif}

In contrast to other lattice-gas models, the Monte-Carlo aspects of
our model allow us to access different scaling regimes by varying
the inverse temperature-like parameter $\beta$ (see Eq.(\ref{eq:bd})
and~\cite{bib:ecb}). 
In lowering $\beta$, the collision step of the time update results in 
a slower phase separation mechanism acting at the interface between 
the two binary fluids, which means that the surface tension is reduced.
Since the bulk viscosity is independent of $\beta$~\cite{bib:cal}, by
lowering $\beta$ we can raise the hydrodynamic length 
$ R_{h} = \frac{\nu^{2}}{\rho \sigma} $~\cite{bib:bray}, where
$\nu$ is the kinematic viscosity, $\rho$ is the density and $\sigma$
is the surface tension coefficient; this enables us to access the high
viscosity regime ($R < R_{h}$).

To analyze the domain growth quantitatively we need to measure a quantity
that captures the length scale of the minority phase domains.
Since our model is symmetric with respect to oil and water particles,
we will call the minority component of our fluid {\it oil}.
As opposed to critical quenches, we now have two typical length scales
in our simulation, one being the size of oil structures, the other
being the one of water structures. Therefore, we cannot now use the
oil-water density difference as input into the pair correlation
function, as done in earlier work~\cite{bib:ecb}. Since we are
interested in the domain growth of the minority phase,
it is convenient to measure the oil based coordinate-space
oil-correlation function. At time $t$ following the quench, this
correlation function is given by
\bge
C({\bf r}, t) = \frac{1}{V} < \sum_{\bf x}{\bf '} (q({\bf x},t) - q_{\it av})
(q({\bf x} + {\bf r}, t) - q_{\it av}) >
\label{eq:cor}
\ee
where $q({\bf x},t)$ is the oil density (total number of oil
particles) at site ${\bf x}$ and time $t$, $q_{\it av}$ is the average
oil density per site, $V$ is the volume, and the average is taken over
an ensemble of initial conditions. 
The prime on the sum implies that the sum extends over all lattice 
sites with $q({\bf x},t) > 0$, i.e. with at least one oil particle per site.
Taking the angular average of $C({\bf r}, t)$ gives $C(r,t)$, 
the first zero crossing of which we take as a measure of the characteristic 
domain size.

All simulations started from random placement of the oil and water particles
on the underlying lattice. Comparative runs have shown that a lattice size of
$256 \times 256$ is large enough to avoid finite-size effects during the
time scales studied. We set $\alpha =1.0$ in Eq.~(\ref{eq:tiw}), and varied 
$\beta$ to access the different fluid regimes as described above. 

Previous studies of critical quenches~\cite{bib:ecb} had shown that,
by using a value of $\beta =0.5$, we are able to access the hydrodynamic
regime  with a growth exponent of $n = \frac{2}{3}$ in critical
quenches. In order to study the behaviour of the growth exponent for a
wide range of concentrations, we kept the reduced density of water $W$
(the majority component) at $W = 0.25$, while varying the reduced
density of oil $O$ from $O =0.04$ up to $O = W$. This translates to
concentrations 
\footnote{Note that the term {\em reduced density}
refers to the ratio of the number of all oil particles in the system
to the number of all possible locations (seven per lattice site) in the
system, which includes {\em empty} locations. The {\em concentration} of
oil, however, is the ratio of the number of oil particles to the
total number of particles.} 
of the minority phase from $C = 13.7\%$ up to $50\%$, increasing in
steps of roughly $2\%$. The behaviour of each system was studied for
five runs over 10000 timesteps each, the domain size $R(t)$ being
measured every 50 timesteps. We plotted $R(t)$ on a double logarithmic
scale to identify possible algebraic growth.

\begin{figure}
\begin{center}
\leavevmode
\hbox{%
\epsfxsize 3.8in
\epsffile{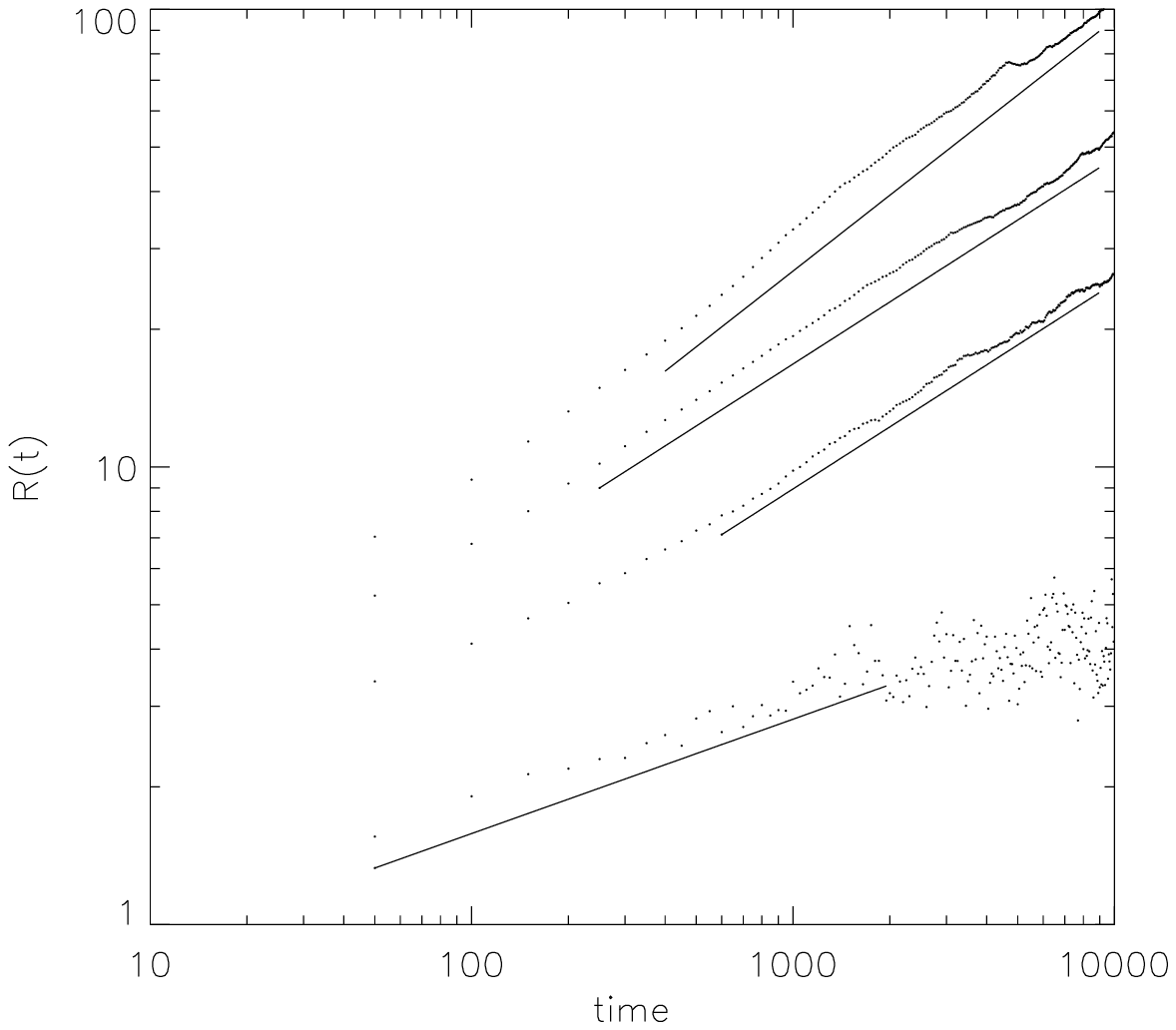}}
\end{center}
\caption{}
\label{fig:cor}
\end{figure}

At very low concentrations of oil, we obtained algebraic growth at
early times and noisy behaviour at late times, which indicates that
the system has reached a point at which the fluctuations dominate over
the phase separation.
Nevertheless, we see a growth exponent of $n = 0.25$ at $C = 13.7\%$,
and a rapid increase in the exponent to $n = 0.40$ at $C = 21.9\%$
as we increase the oil concentration. With increasing oil concentration we
observe that the noise reduces and vanishes for oil concentrations above
$20\%$. We explicitly checked for finite size effects in this region
and found that the domain growth exponent measured on a $512 \times
512$ lattice does not differ from our results from the smaller systems.
Starting from concentrations of $C = 24\%$ up to $38\%$ we find that
the domains grow algebraically in time with an exponent $n = 0.45$.
This plateau in the growth exponent indicates the existence of a
scaling regime in the late time kinetics, which are dominated by
hydrodynamic effects. Similar values have been
reported by other authors~\cite{bib:veto,bib:cn,bib:walc}, but
our results allow us to exclude a smooth transition between $n =
\frac{1}{3}$ and $n = \frac{2}{3}$ as proposed by one of these
papers~\cite{bib:veto}. The measured growth exponent contradicts,
however, the results of Osborn {\em et al.}~\cite{bib:oosyb}, who
reported $n = 0.56\pm0.03$. We believe that the reason for this
discrepancy is precisely the same as in the case of critical quenches,
namely the absence of fluctuations in their model, which seem to play
a vital role for early time and off-critical quenches.  
This result is also below the theoretical prediction by
Furukawa~\cite{bib:f2} and San Miguel {\em et al.}~\cite{bib:smgg}. We
will discuss possible reasons for this at the end of this section.
As the concentration of oil is increased further, the growth exponent
increases quickly with concentration and finally reaches its value from
the critical quench of $n = \frac{2}{3}$ at concentrations of $47\%$ onwards.
For typical results of this first set of simulations compare
Fig.~\ref{fig:cor}.

\begin{figure}
\begin{center}
\leavevmode
\hbox{%
\epsfxsize 3.8in
\epsffile{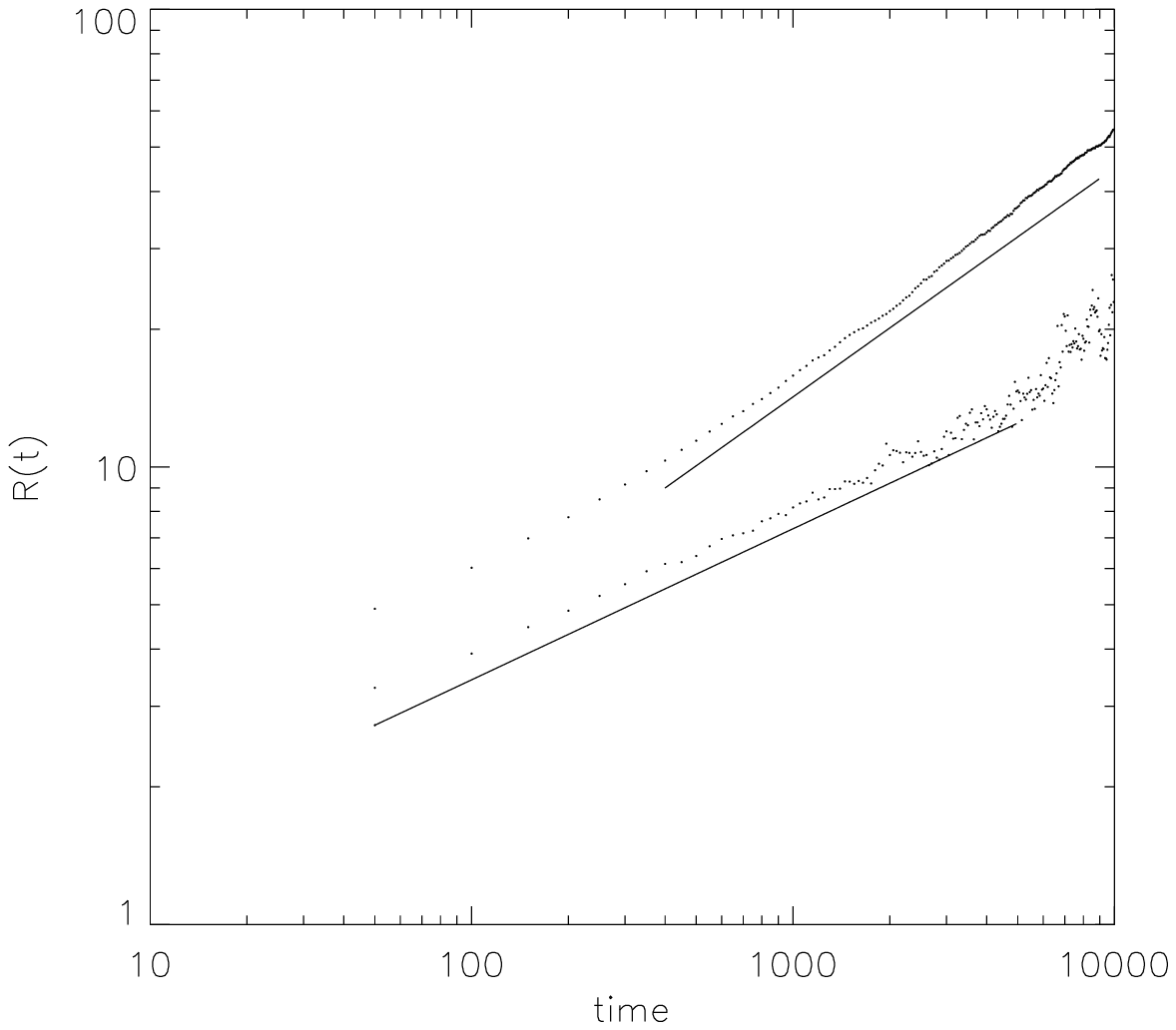}}
\end{center}
\caption{}
\label{fig:visc}
\end{figure}

\begin{figure}
\begin{center}
\leavevmode
\hbox{%
\epsfxsize 3.8in
\epsffile{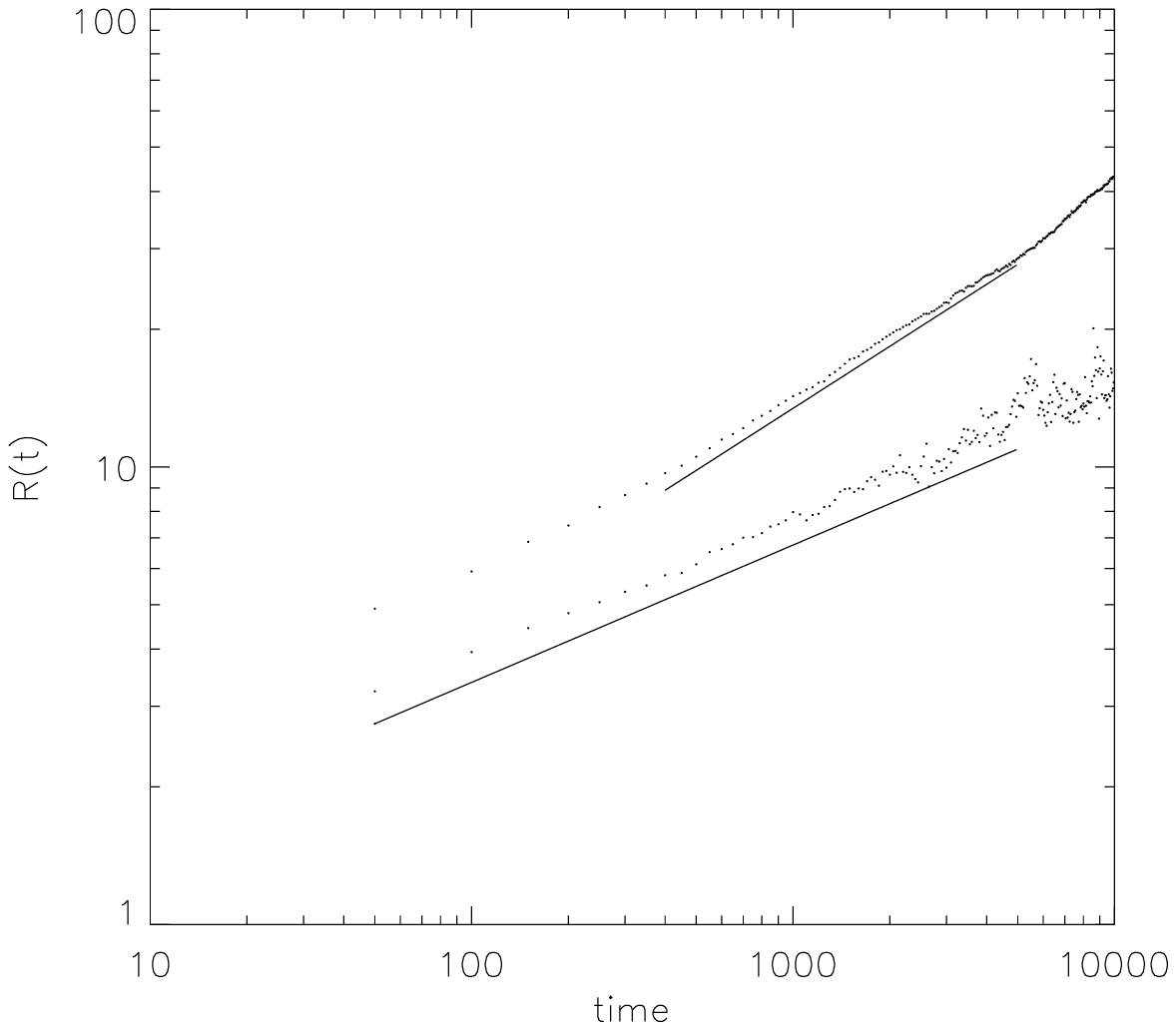}}
\end{center}
\caption{}
\label{fig:real}
\end{figure}

In order to study the early time regime in more detail, we performed
additional simulations at lower values of $\beta$, where we
observe the early time growth in the critical quenches. As described
above, a lowering of $\beta$ corresponds to an increase in the
hydrodynamic length $R_h$.
In this case we chose $\beta = 0.15$, which leads to a crossover between
the viscous and the hydrodynamic regime in the critical quench. A
further decrease to $\beta = 0.137$ results in domain growth with
$n = \frac{1}{2}$~\cite{bib:ecb} throughout the observed time scales,
which is evidence for the viscous regime. 
The decrease in $\beta$ means, however, that we also raise the
effective temperature in our system, thus increasing fluctuations.
These fluctuations again become dominant at small concentrations of
oil. Despite this effect, we are able to get reliable results from
concentrations as low as $30\%$. From concentrations of $40\%$
onwards, the noise vanishes from our simulations, and we observe clear
growth exponents of $n = 0.3$ and $n = 0.33$ for $\beta = 0.137$ and
$\beta = 0.15$, respectively. Some results are displayed in
Fig.~\ref{fig:visc} for $\beta =0.15$ and Fig.~\ref{fig:real}
for $\beta = 0.137$. 
Unlike the late time dynamics, we do not find a unique universal growth
exponent over a wide range of concentrations. The strong dependence of
$n$ on the minority phase concentration could explain
why several authors~\cite{bib:veto,bib:oosyb,bib:cl,bib:veto2} have reported
values considerably lower than the value of $n \approx 0.45$ found in most
studies. These anomalies may be due to the fact that within such
simulations only the early time regime, for which behaviour
depends on the concentration and on the temperature, is probed. 
This might be evidence that we do not have a single, dominant
scaling regime in the early time droplet growth in off-critical quenches.
In this case the question remains as to what the other growth
regimes are. One possible candidate is certainly the Lifshitz-Slyozov
evaporation-condensation growth with $n = \frac{1}{3}$, but our
results do not show any evidence for this. 
In a renormalisation group analysis, Corberi {\em et
al.}~\cite{bib:ccz} recently speculated about another very early time
regime in binary immiscible fluid phase segregation, which might be
due to a newly discovered fixed point. Their Langevin approach lead
them to an exponent of $n = \frac{1}{4}$. This is exactly the lowest
value that we find for $\beta = 0.5$ as seen in Fig.~\ref{fig:cor}.
Our results in the simulations with $\beta = 0.15$ and
$\beta = 0.137$, however, do not agree with their conclusions, since
our lowest measurable growth exponent was $n = 0.2$, but as mentioned above,
these simulations were perturbed by strong fluctuations. At this
parameter setup we performed additional simulations on a $512 \times 512$
lattice to exclude finite size effects and get excellent agreement
between the different runs. We also note that molecular dynamics simulations
undertaken by Velasco and Toxvaerd~\cite{bib:veto} found a very slow
growth regime as well. Results from our ternary amphiphilic fluid studies
even indicate that there is no lower threshold for $n$ (see discussion in
Sec.~\ref{sec:tern} \& ~\ref{sec:dac}).

A different explanation for the variance of the early
time exponents might be that corrections to a scaling regime lead to
the measurement of exponents which are time dependent and only
asymptotically reach the predicted values. 
This mechanism was indeed responsible for the non-observance of
Lifshitz-Slyozov growth in Monte-Carlo simulations of an Ising
model~\cite{bib:huse}. In fact, the prediction of Yao {\it et
al.}~\cite{bib:yegg} of $R(t)^3 = A + Bt$ in non-hydrodynamic
off-critical quenches would provide an argument for this explanation.
By plotting $R^3$ against $t$ we tried to detect this growth
law, but did not find evidence for it. 
A detailed study of this most interesting question would involve the
measurement of effective exponents and require extensive computational
effort which is beyond the scope of this paper.

Another interesting fact is that a convincing explanation of the
shift $\delta \approx 0.05$ in the late time growth exponent
$n = \frac{1}{2} - \delta$ is still missing. All except one simulation
report results below the value of $n = \frac{1}{2}$ derived by
Furukawa~\cite{bib:f2} and San Miguel {\em et al.}~\cite{bib:smgg}.
Furukawa himself mentions possible deviations from this growth
exponent due to internal flows among droplets caused by the surface
tension, which cannot be neglected in the off-critical quench. The
measured exponents therefore may still be in the
time-dependent crossover regime, which is anomalously -- or according
to Furukawa even infinitely -- elongated. To understand the mechanism
behind the domain growth law, simulations in three dimensions are
highly desirable. In three dimensions the predictions by Furukawa and
San Miguel {\em et al.} differ, so simulations and experiments should
be able to determine the growth regime.  

\begin{figure}
\begin{center}
\leavevmode
\hbox{%
\epsfxsize 3.8in
\epsffile{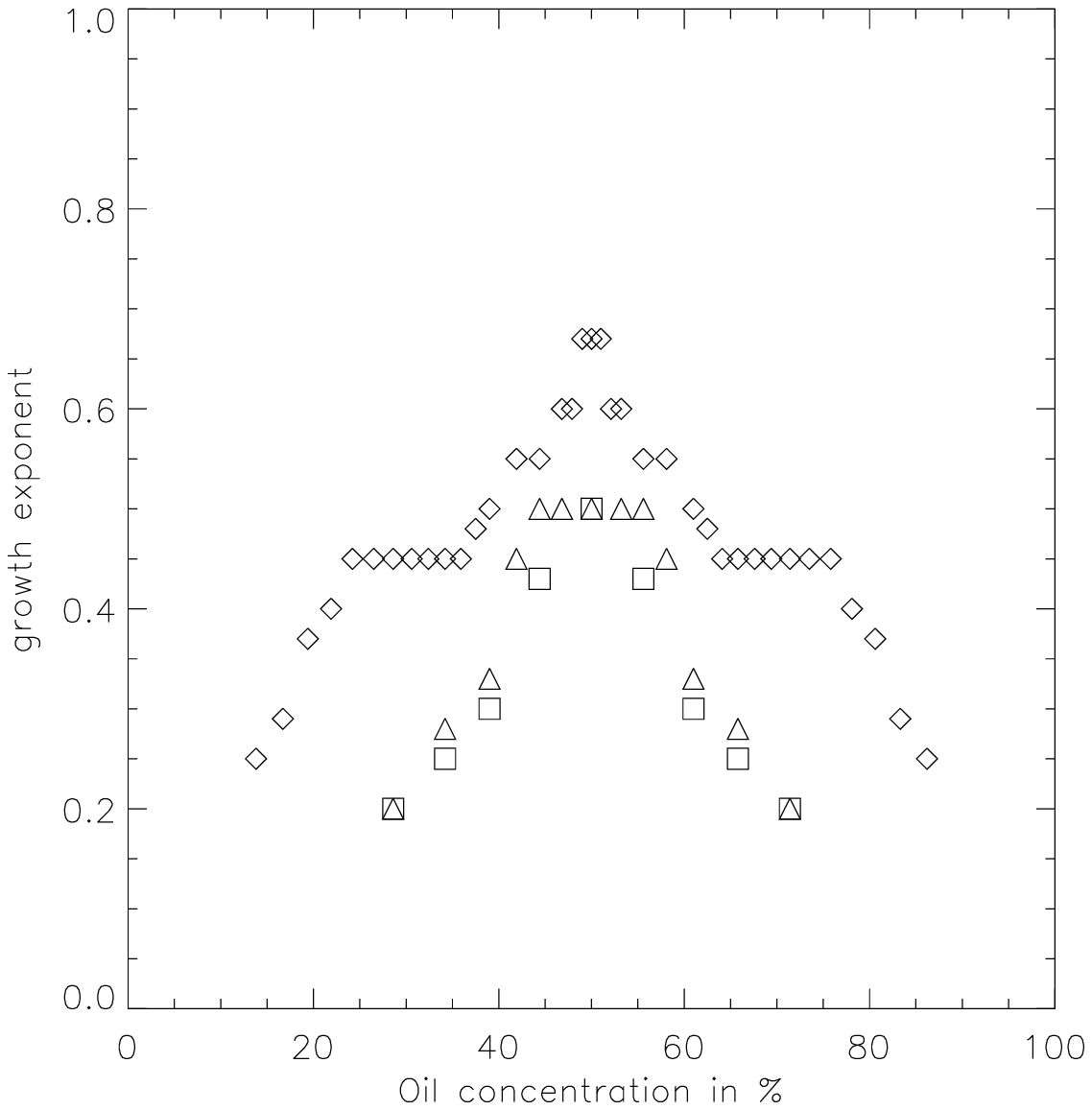}}
\end{center}
\caption{}
\label{fig:comp}
\end{figure}

A summary of all our results from the binary fluid simulations is
given in Fig.~\ref{fig:comp}, which shows how the growth exponent
depends on the concentration of oil for the different regimes.

\section{Domain growth in ternary amphiphilic fluids}
\label{sec:tern}

We now turn to the analysis of the ternary amphiphilic system. We
expect that the presence of surfactant in an oil-water mixture will
change the domain growth behaviour dramatically, due to the reduction
of surface tension between oil-water interfaces. Starting from a
binary mixture of oil and water,
where oil once again is the minority phase, adding small amounts of
surfactant will not change the growth kinetics on the time scales
observed. With increasing concentration of amphiphile, however, the drop in
surface tension will take the system to progressively earlier growth
regimes. From our results in the binary immiscible fluid case above,
we therefore expect an algebraic growth regime with growth exponent $n
= 0.45 $ at low surfactant concentrations. We expect to see that an
increase in the reduced density of surfactant, $S$, leads to a
decrease in the growth exponent, at least up to $n = 0.20$, our last
reliable result in the binary fluid case. In the presence of
amphiphile we are now able to access early time regimes without
lowering $\beta$; therefore we do not have to deal with noise as in
the binary fluid regime, so we might actually be able to speculate
about very early time regimes. Of course, one important
{\it caveat} has to be mentioned here: the presence of surfactant molecules
may have various effects on the domain growth, so we must be cautious
in extrapolating the ternary results to the binary immiscible fluid system.

Moreover, with sufficient surfactant in our system, we should see the 
formation of a stable {\it droplet phase}, in which oil particles will be 
surrounded by a monolayer of surfactant, that separates the oil from the 
water domains. The fact that surfactant particles have to sit at
oil-water interfaces, must eventually lead to arrest of domain growth.
The oil droplets reach an average droplet size $R_d$ that
we expect to be inversely proportional to the amount of surfactant
present at the oil-water interfaces~\cite{bib:ecb}. 

Previous studies~\cite{bib:ecb} of self-assembly kinetics in the
sponge phase showed a crossover to logarithmically slow growth with
increasing surfactant concentration. At sufficient high surfactant
concentration, the average domain size was fitted phenomenologically
to a stretched-exponential form,

\bge
R(t) = A - B \exp (-Ct^D) ,
\label{eq:expo}
\ee

\noindent in which the parameters $A, B, C$ and $D$ have to be determined
numerically. 

For our simulations we used a $128\times128$ lattice, starting from a
random placement of the particles on the lattice. We set the
temperature-like parameter to $\beta = 1.0$
in order to reduce noise in our simulations, but without slowing down the
kinetics. For the coupling constants in the interaction energy of
Eq.(\ref{eq:tiw}) we used

\bgeas
	\alpha   &=& 1.0   \\
	\mu      &=& 0.5   \\
	\epsilon &=& 8.0   \\
	\zeta    &=& 0.05 
\eeas 

\noindent which will encourage the amphiphilic dipoles to sit at oil-water
interfaces, as well as bending of amphiphilic interfaces, both
characteristic features of the droplet phase~\cite{bib:bce}.
In our simulations, we kept the water:oil ratio fixed,
while varying the concentration of surfactant. We studied
water:surfactant:oil ratios from $25:2:8$ up to $25:30:8$, while
keeping the total reduced density between $0.35$ and $0.45$. 
We will describe our results in terms of the surfactant:oil ratio
$\Gamma = \frac{S}{O}$.
All simulations were ensemble-averaged over at least five runs.
The average oil droplet size was again taken to be the first zero of
the correlation function defined in Eq(\ref{eq:cor}).

\begin{figure}
\begin{center}
\leavevmode
\hbox{%
\epsfxsize 3.8in
\epsffile{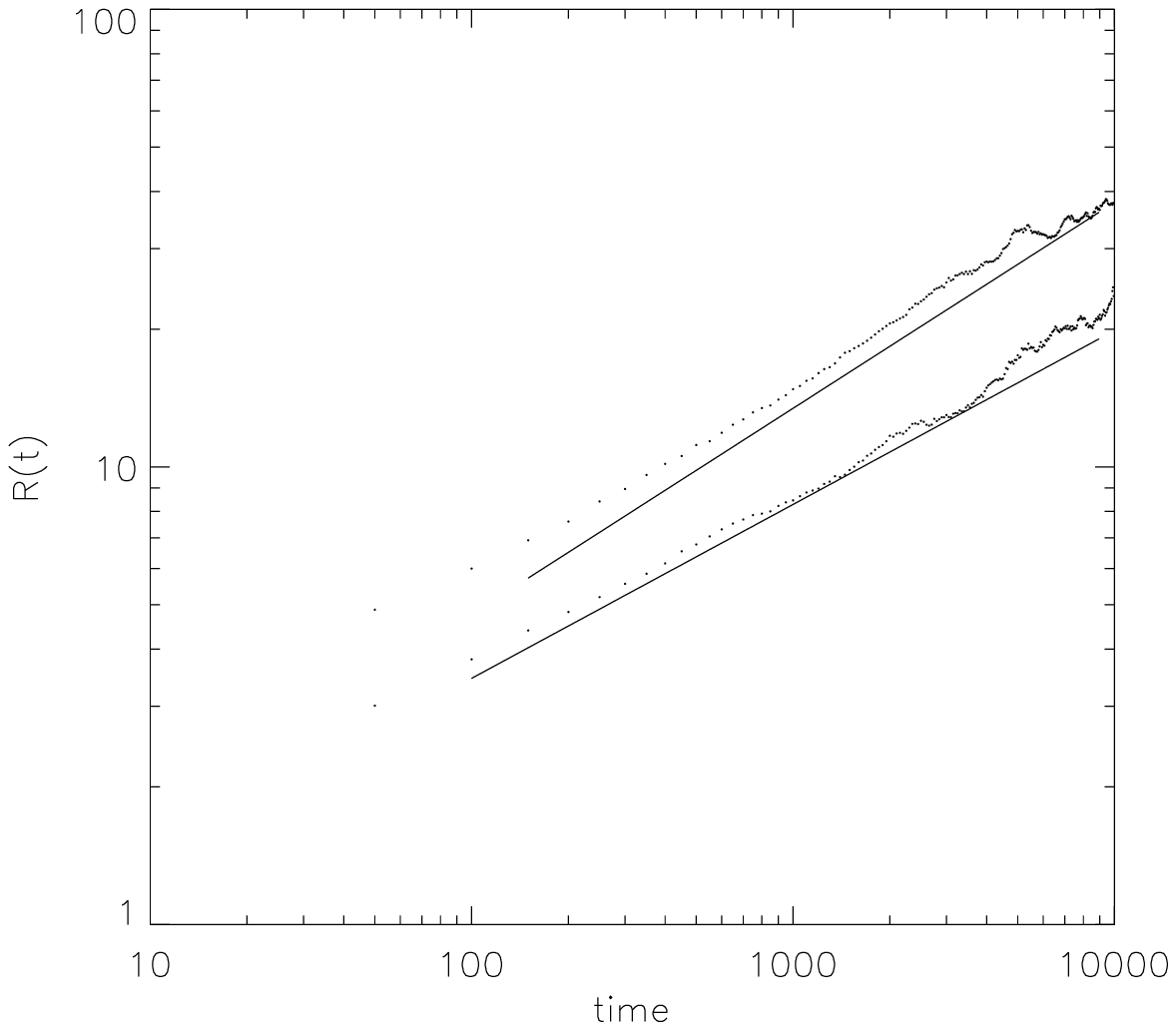}}
\end{center}
\caption{}
\label{fig:tlin}
\end{figure}

The results for systems with $\Gamma = 0.25$ and $\Gamma = 0.625$ are
shown in Fig.~\ref{fig:tlin}. In both cases we see algebraic
growth throughout the time scale studied. Whereas for $\Gamma = 0.25$ the
growth exponent was found to be $n = 0.45$ as in the corresponding off-critical
quench in the binary fluid system, we observe a slight decrease in the
exponent for $\Gamma = 0.625$, where $n = 0.38 $. This is due to the
fact that the presence of surfactant reduces the oil-water
interfacial tension, therefore reducing the driving force of the phase
separation. Similar behaviour has been found for values of $\Gamma$ between
the two shown here. Visualization of these simulations demonstrated
the formation of oil bubbles which are partially covered by a thin
layer of surfactant. In all these simulations we did not see
any domain saturation effects due to the presence of surfactant. 

\begin{figure}
\begin{center}
\leavevmode
\hbox{%
\epsfxsize 3.8in
\epsffile{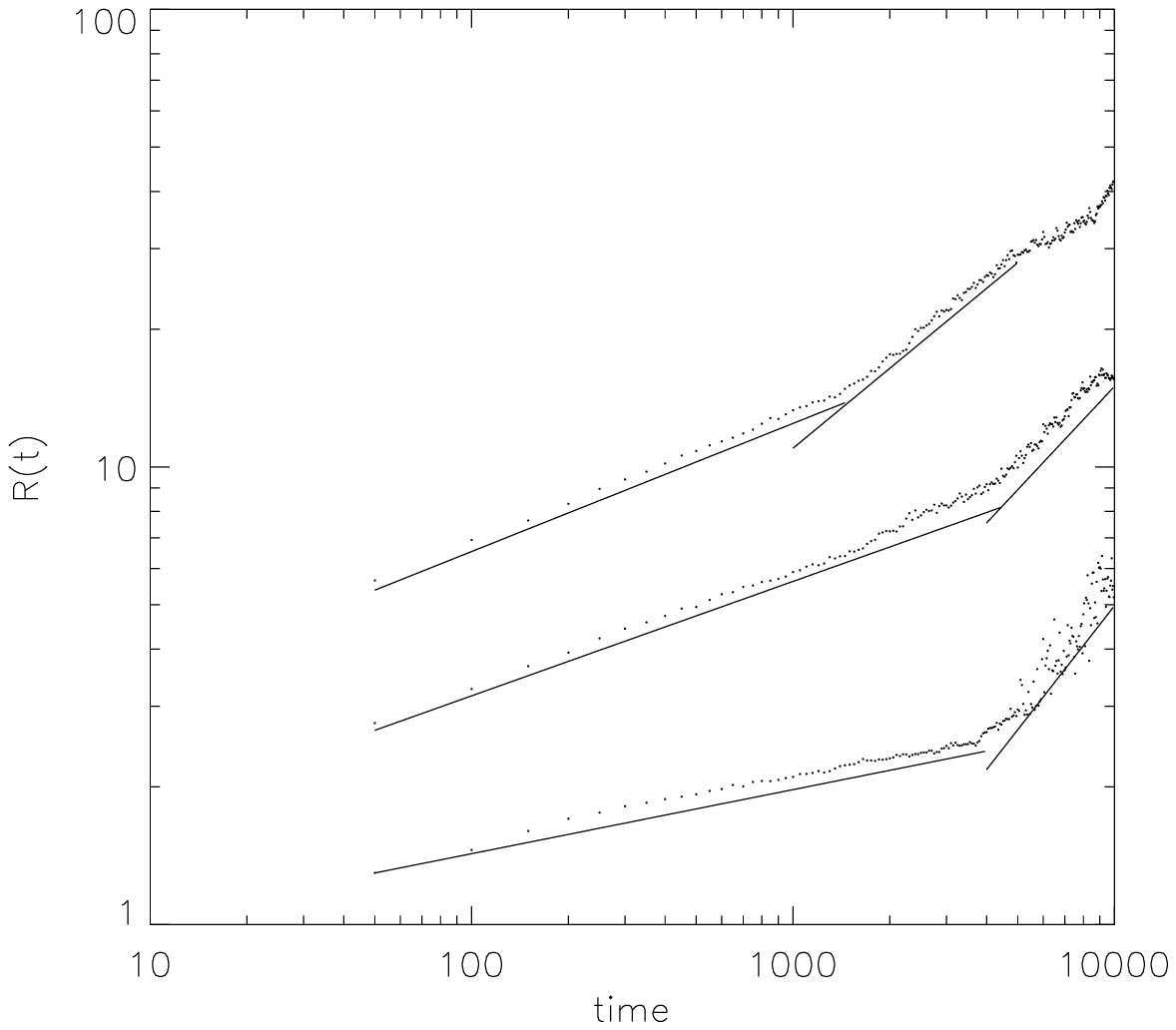}}
\end{center}
\caption{}
\label{fig:tnuc}
\end{figure}

At $\Gamma = 0.75$ we observe a slight kink in the domain size {\it
vs.} time plots, indicating faster algebraic growth at late times.
First we expected this to be due to the crossover to the regime
with $n = 0.45$. A further increase to $\Gamma = 1$ showed that we
were dealing with a new kind of behaviour, since the growth exponent
in the late time regime was clearly above the expected value. 
Plots with $\Gamma = 1.175, 1.375$ and $2.0$ are given in 
Fig.~\ref{fig:tnuc}. We observe this phenomenon over a wide
range of surfactant concentrations, starting from $\Gamma = 0.75$ up
to $\Gamma = 2.0$. We explicitly checked that the kink is not due
to a finite size effect. The results from five simulations with
$\Gamma = 1.5$ on a $256 \times 256$ lattice are in excellent
agreement with the results found on the usual $128 \times 128$ lattice.  
A summary of all our simulations with algebraic early-time growth
showing this kind of behaviour is given in Table~\ref{tab:nuc}.
 
\begin{figure}
\begin{center}
\leavevmode
\hbox{%
\epsfxsize 6.0in
\epsffile{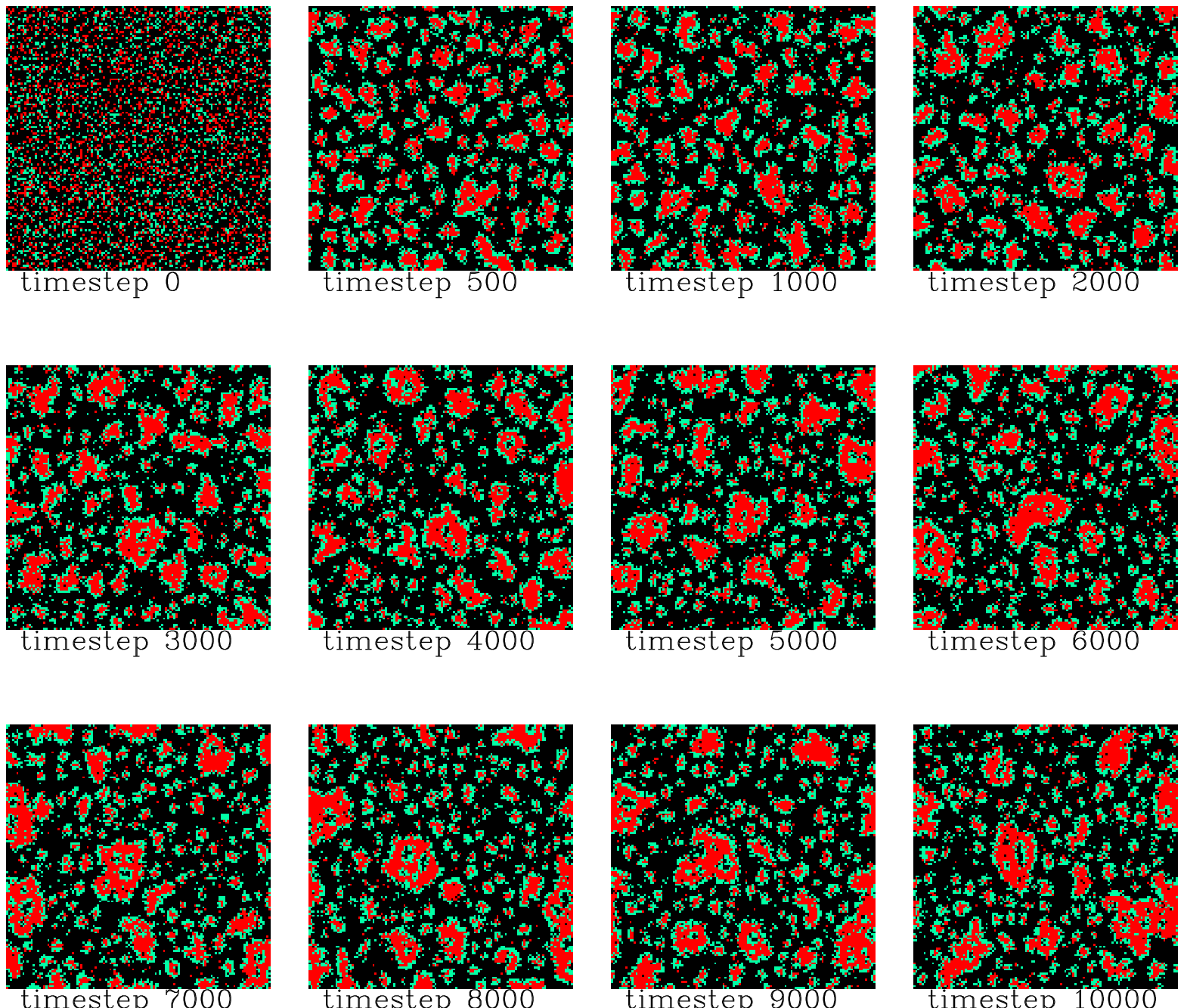}}
\end{center}
\caption{}
\label{fig:tvnuc}
\end{figure}

To examine the physical significance of this process, we visualized
system configurations for several simulations.
A typical result is shown in Fig.~\ref{fig:tvnuc}, where $\Gamma =
1.5$. At this surfactant concentration, the faster late time growth
starts after $3000$ timesteps. We explain this phenomenon by a
{\em nucleation process}.
While some droplets above a possible critical size $R_c$ continue to
grow, smaller droplets in the surrounding areas shrink; therefore our
system now contains two length scales. Moreover, as is clearly
visible in the figure, the larger oil droplets contain water droplets
of size similar to the smaller oil droplets. 
This is evidence that our system has moved from the oil droplet phase
as seen at $T = 1000$ timesteps into a coexistence region of oil
and water droplet phases, in which two length scales exist, one
connected to the size of all small droplets and one corresponding to
the size of the larger oil droplets.  The transition between a pure
phase and two coexisting phases may be connected with an energy
threshold, which explains the nucleation process that we see in our
simulations.

\begin{figure}
\begin{center}
\leavevmode
\hbox{%
\epsfxsize 3.8in
\epsffile{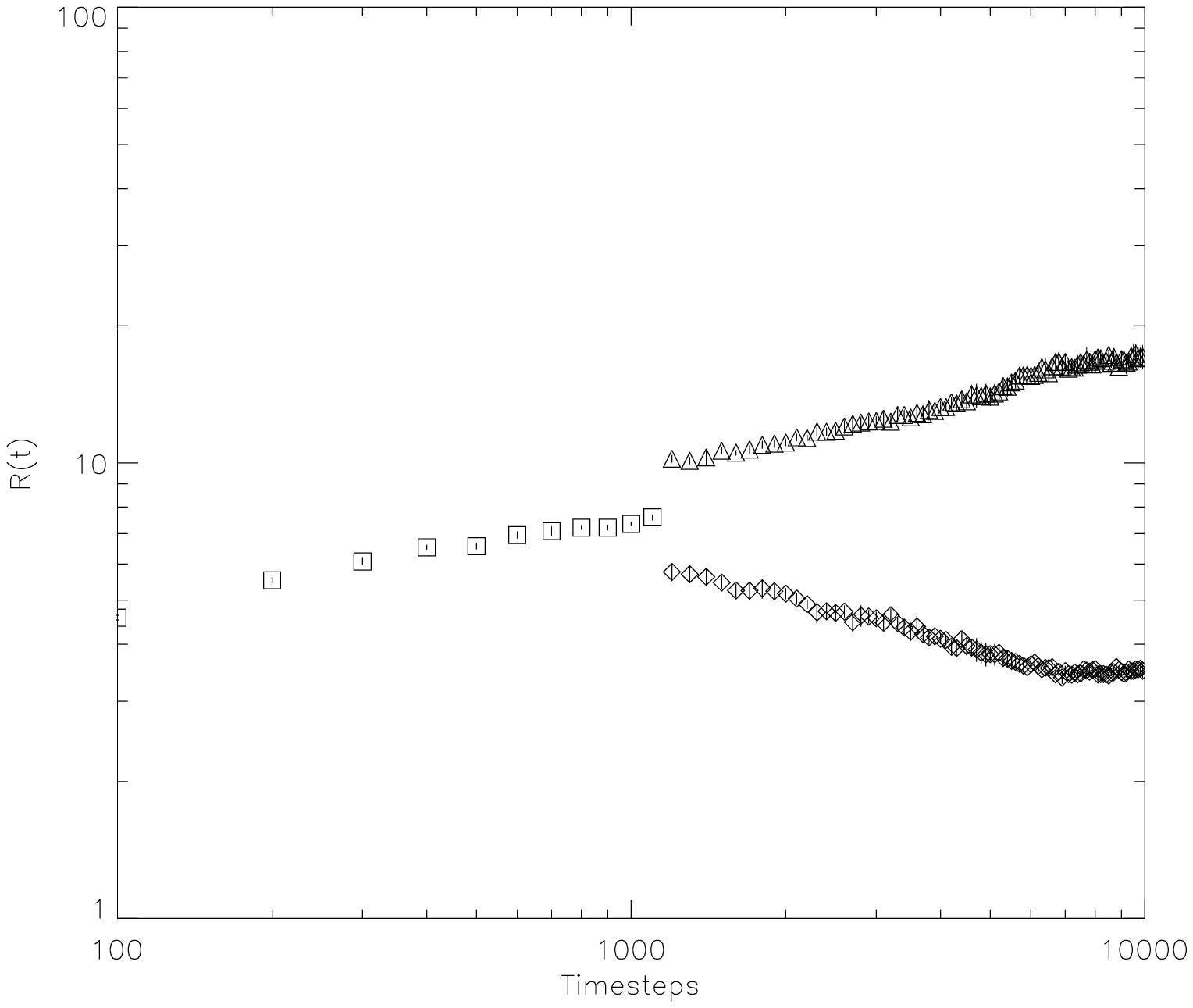}}
\end{center}
\caption{}
\label{fig:twoscales}
\end{figure}

In order to get quantitative evidence for this, we measured the sizes
of droplets directly from the
visualized data and plotted a histogram of number of droplets
{\it vs.} droplet area for several timesteps. We observe that,
associated with the nucleation process, a gap emerges in the
histogram. Whereas at early times about $75\%$ of all droplets have
radii between $5.5$ and $11$, this value drops below $10\%$ at
timesteps after nucleation. The existence of this gap can be shown
very drastically: Although we allow a large overlap of `small'
(radii between 2 and 11) and `large' (radii between 5.5 and 50)
droplets, Fig.~\ref{fig:twoscales} shows
that we are really dealing with two length scales in our system. These
plots of the time evolution of small and large droplets show opposite
behaviour after a common early-time growth: Whereas the size of the smaller
droplets decreases slightly during the time scales observed, the
larger droplets grow considerably. Similar studies have been made for
several values of $\Gamma$, all showing the same typical behaviour.
However, the characteristics of the existence of two length scales
become weaker with increasing surfactant concentration. This is
connected to the noisy behaviour in the fast growing late-time regime,
which is visible in Fig.~\ref{fig:tnuc} for $\Gamma = 2.0$. We explain
this as follows: Since an increasing amount of surfactant means that the
average size of a droplet shrinks, the distinction between larger and
smaller droplets becomes more difficult to make, and even after the
nucleation takes place, there is a rapid exchange of particles
between the two typical droplet sizes. Therefore the gap in the
droplet size histogram vanishes gradually. As a result we see a wildly
fluctuating overall domain size after the nucleation takes place.

Perhaps most interesting is that the surfactant molecules seem to
play two distinct and opposing roles in the two growth regimes.
The early time growth is further and further suppressed by
the presence of amphiphiles in our system, resulting in
a drop of the growth exponent from values of $n= 0.3$ at $\Gamma = 1.125$ to 
$n = 0.14$ at $\Gamma = 2.0$. This can be explained by the reduction of
oil-water interface tension, resulting in a decrease of the phase
separation driving force. After the nucleation, however, the
growth exponent increases with $\Gamma$ from values of $n = 0.58$ at
$ \Gamma = 1.125$ up to $n = 0.9$ at $\Gamma = 2.0$. Therefore we
conclude that surfactant molecules play a catalytic role in the
nucleation process.  

\begin{figure}
\begin{center}
\leavevmode
\hbox{%
\epsfxsize 3.8in
\epsffile{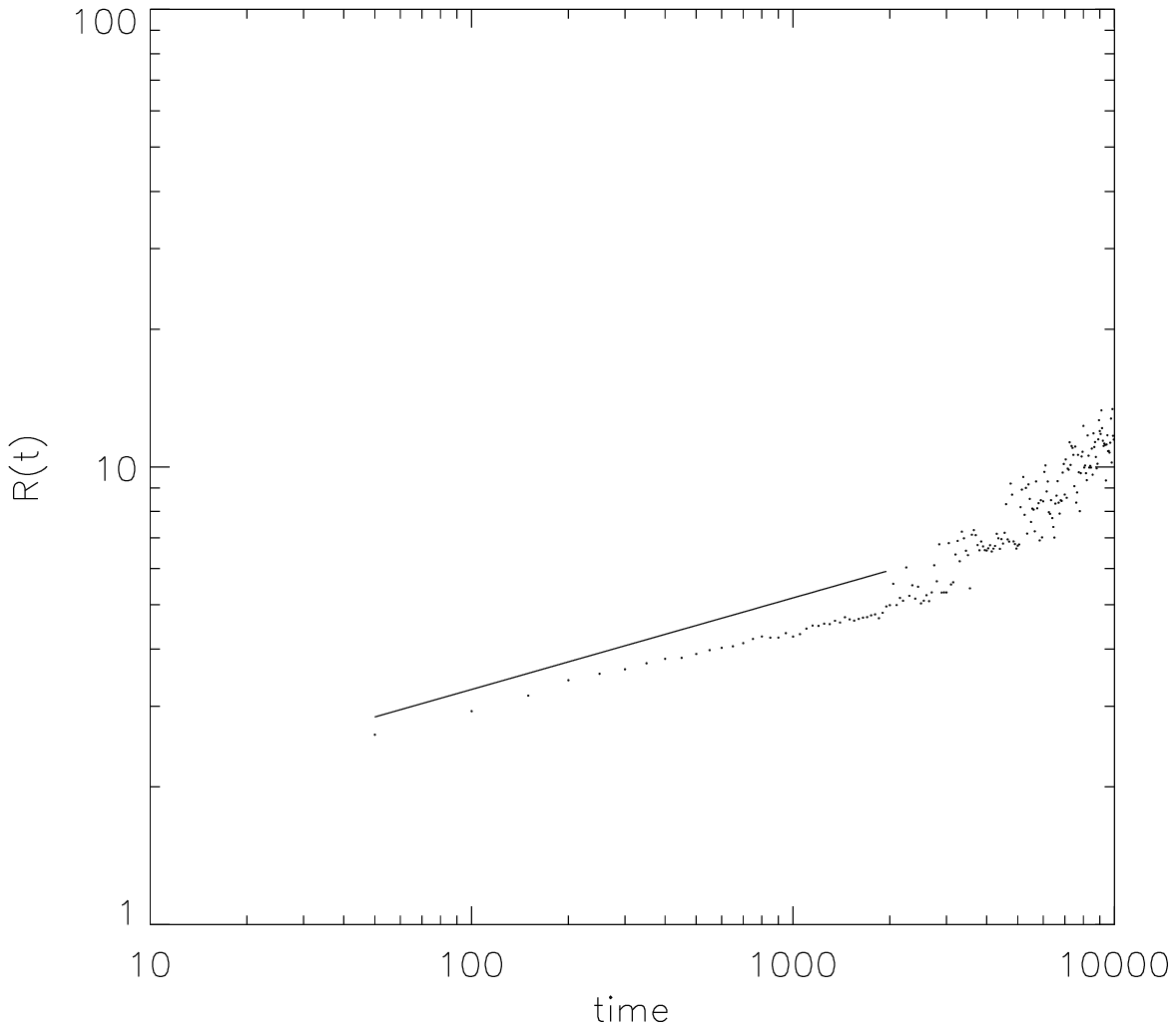}}
\end{center}
\caption{}
\label{fig:t19}
\end{figure}

\begin{figure}
\begin{center}
\leavevmode
\hbox{%
\epsfxsize 3.8in
\epsffile{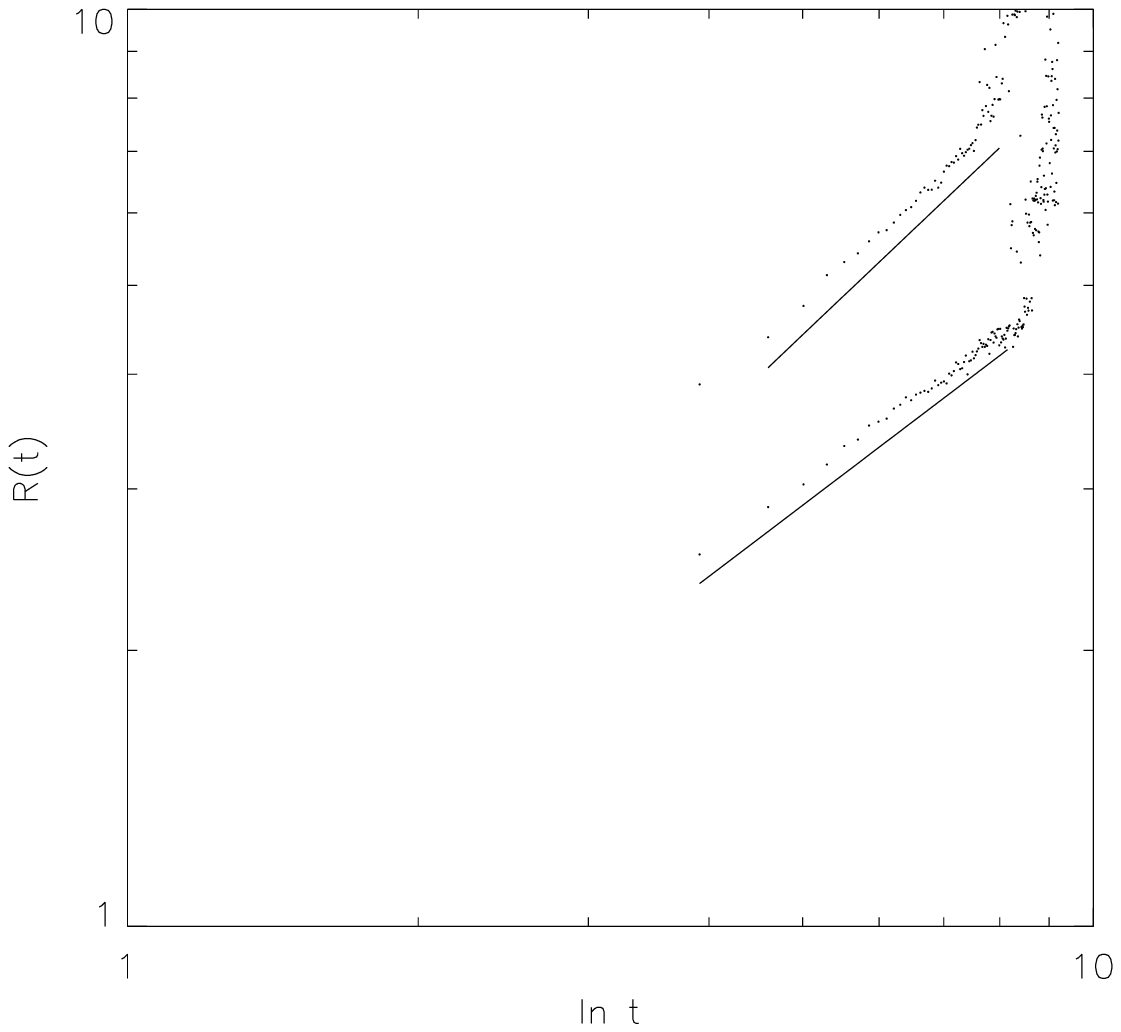}}
\end{center}
\caption{}
\label{fig:tlog}
\end{figure}

When we increase the surfactant concentration still further, we
observe deviations from the algebraic early-time growth, as shown in
Fig.~\ref{fig:t19}, whereas the late time behaviour (in spite of the
noise mentioned above) can still be described as algebraic growth.
In order to investigate whether the presence of surfactant now leads to
logarithmically slow growth, we have plotted the domain size {\it vs.}
$\ln t$ on a logarithmic scale. The results for $\Gamma = 2.375$
and $\Gamma = 2.875$ are shown in Fig.~\ref{fig:tlog}. This kind of
logarithmic growth, $R(t) \propto (\ln t)^{\theta}$, suggests that our
system shares some common features with pinned domain growth in
systems with quenched disorder~\cite{bib:bray}, reported by Laradji et
al.~\cite{bib:lmtz} and in the study of the self-assembly kinetics of
the sponge phase using the same lattice gas model~\cite{bib:ecb}.
We observe logarithmic growth in the early time regime
from a surfactant:oil ratio $\Gamma = 2.175$ up to $\Gamma = 3.0$,
during which the exponent $\theta$ decreases from $1.05$ to $0.7$. This slow
growth of the oil droplets indicates that we are now very close to the
point when the oil droplet domains become saturated. At late times,
however, we still observe nucleation followed by very noisy growth.

\begin{figure}
\begin{center}
\leavevmode
\hbox{%
\epsfxsize 3.8in
\epsffile{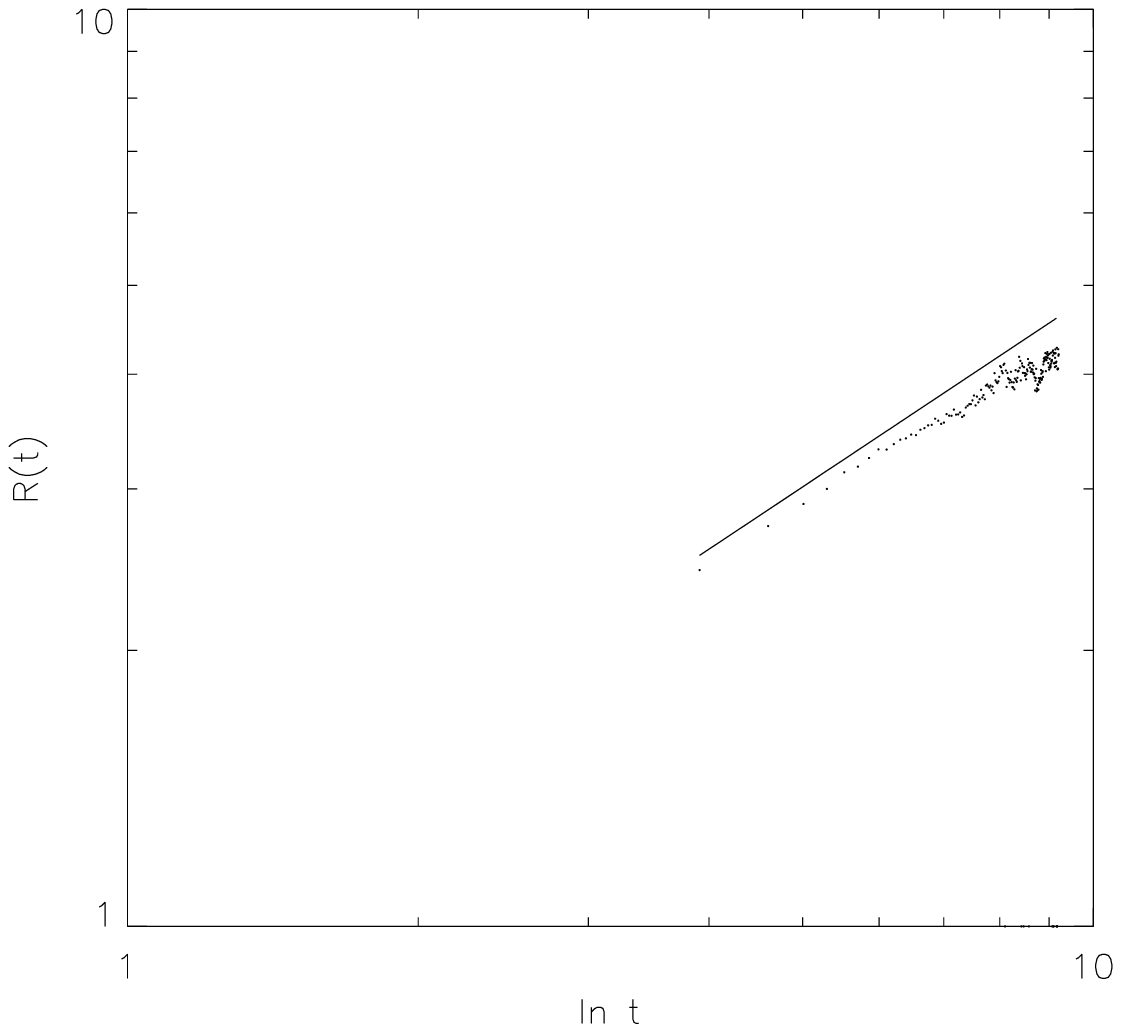}}
\end{center}
\caption{}
\label{fig:t24}
\end{figure}

\begin{figure}
\begin{center}
\leavevmode
\hbox{%
\epsfxsize 3.8in
\epsffile{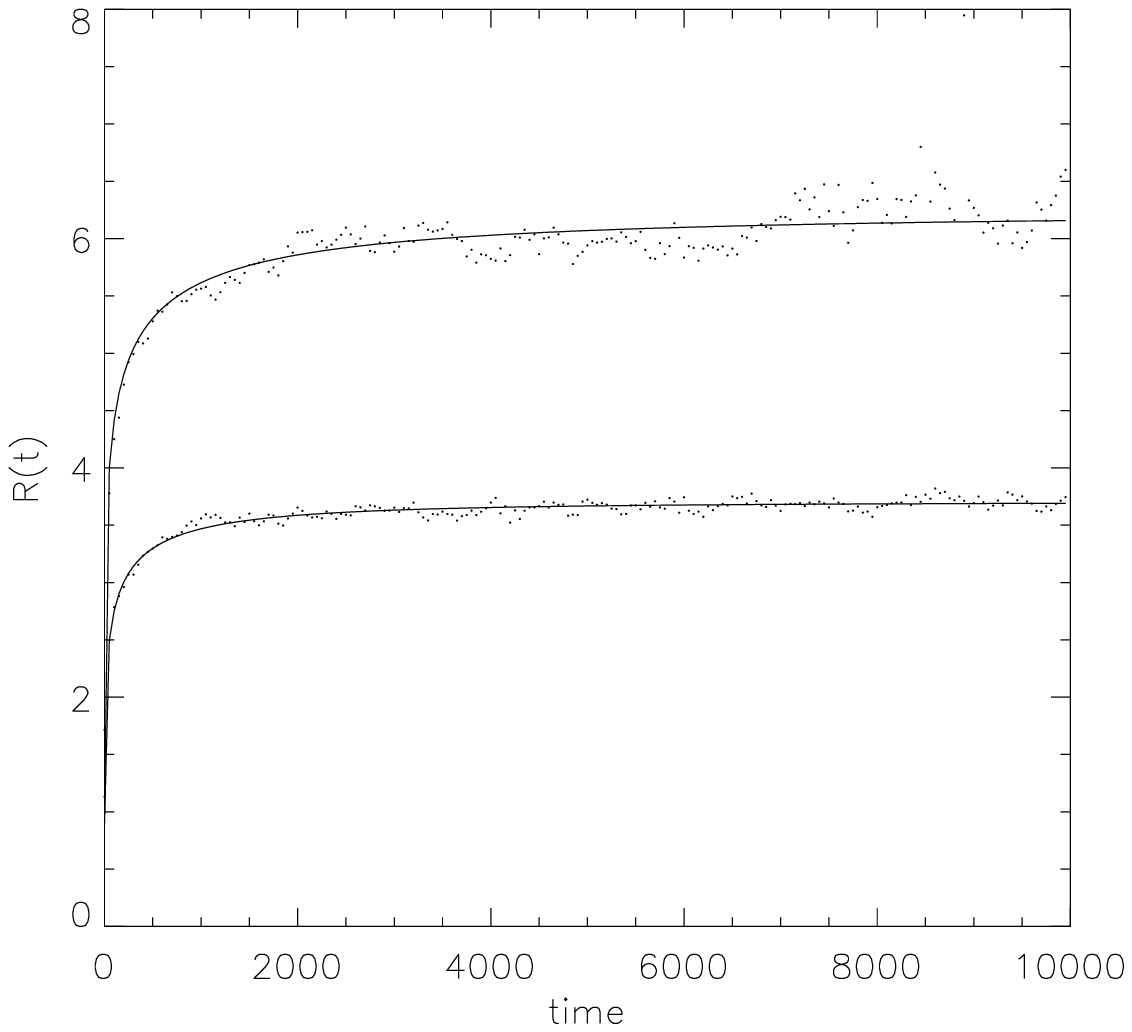}}
\end{center}
\caption{}
\label{fig:tsat}
\end{figure}

Saturation of oil droplets is first reached at $\Gamma = 3.0$, when we see
clear deviations from the logarithmic growth at $t > 3000$ timesteps, as
shown in Fig.~\ref{fig:t24}. Also the nucleation process stops, since
the noise at late times vanishes --- the tendency of surfactant to sit at
oil-water interfaces has overcome the immiscibility of oil and water.
As shown in Fig.~\ref{fig:tsat}, oscillations in the average droplet
size decrease with a further increase of the surfactant
concentration. These fluctuations also occurred in the study of the sponge
phase~\cite{bib:ecb}, and correspond to the continual break up and
reformation of droplets, resulting from the finely balanced
competition between the immiscibility of oil and water and the action
of the surfactant molecules. We again checked explicitly whether finite
size effects lead to arrested domain growth and found none on the
time scales of our simulations. The fit of domain growth to the
stretched-exponential form (Eq.\ref{eq:expo}) is also included in this
figure and successfully describes the behaviour of the system over the
whole time-scale of these simulations. Of the four coefficients of
this function, only the decay rate $C$ and the stretching exponent $D$
are of physical interest. Whereas $C$ does not depend very strongly on
the surfactant concentration, $D$ increases with it. 

A possible explanation of this stretched exponential form of domain
growth can be given, when we assume that close to the saturation
point relaxation times $\tau = \frac{1}{\lambda}$ accumulate in our
system. If we also include an essential singularity
$\exp{(-a \lambda^{-k})}$ as a weighting function and cutoff of the
different relaxation times, we then get 

\bge
R(t) \propto \int d\lambda \:\: \exp{(-(\lambda t + \frac{a}{\lambda^k}))}
\label{eq:expexp}
\ee

\noindent instead of a simple exponential decay. Applying the method
of steepest descent to the integrand, we derive $R(t) \propto
\exp{(- Ct^D)}$, where $C \equiv (k+1)
\left(\frac{a}{k^k}\right)^{\frac{1}{k+1}}$ and $D \equiv
\frac{k}{k+1}$. If we look at the values of $a$ and $k$ as functions
of the surfactant concentration, we see that both go asymptotically
to a finite value that is always reached at $\Gamma =
3.375$. Whereas $a$ decreases to approach $0.03$, $k$ increases until
it reaches $0.57$. This asymptotic behaviour might indicate that $a$
and $k$ are more fundamental variables than $C$ and $D$.
Stretched exponential decay has been reported by NMR experiments in
porous media~\cite{bib:kenyon} and it seems plausible that the complex
systems of surfactant mixtures inhibit similar features. 
However, we are far from an understanding of the physical
reasons of the accumulation of relaxation times.

\section{Transition between Droplet and Sponge Phase}
\label{sec:trans}

In addition to our other results, we also performed several
simulations to study the transition between the droplet and the sponge
phase. In order to access both of these different phases, we changed
our interaction parameters to the set used in~\cite{bib:bce}, which
already proved to accommodate both phases, depending on the concentrations
of oil, water and surfactant. In our simulations we used a $128\times128$
lattice, set the temperature-like parameter $\beta = 1.0$, and the
interaction parameters to

\bgeas
	\alpha   &=& 1.0   \\
	\mu      &=& 0.05   \\
	\epsilon &=& 8.0   \\
	\zeta    &=& 0.5 .
\eeas 

\noindent There are some slight differences in these values compared
to those used in Sec.~\ref{sec:tern}. We increased $\zeta$ and
decreased $\mu$ in order to reduce the tendency of the interfaces to
bend, while promoting the propensity of surfactant to align
at oil-water interfaces. We kept the reduced
densities of water and surfactant fixed at $W = 0.18$ and $S = 0.15$,
while varying the reduced density of oil from $O = 0.04$ up to $O =
0.18$. This means that we start with a dilute oil phase, where the
surfactant-oil ratio is $\Gamma = 3.75$, and so well above the value
we found necessary in the previous section to form a stable droplet
phase. The upper limit with equal amounts of oil and water still
contains enough surfactant to form a sponge phase~\cite{bib:ecb}. 

We expect that with increasing oil concentration the saturated state
would contain increasingly larger and deformed droplets indicating the
tendency to reduce the interfacial free energy and the consequential
formation of larger structures. At one point, these deformed droplets
will percolate, resulting in the sponge phase. An interesting question
is whether the stretched-exponential form of domain growth can be
consistently applied during the phase transition. Whereas this form was
shown to give good results in both limits, the applicability in
the crossover has yet to be shown. 

\begin{figure}
\begin{center}
\leavevmode
\hbox{%
\epsfxsize 3.8in
\epsffile{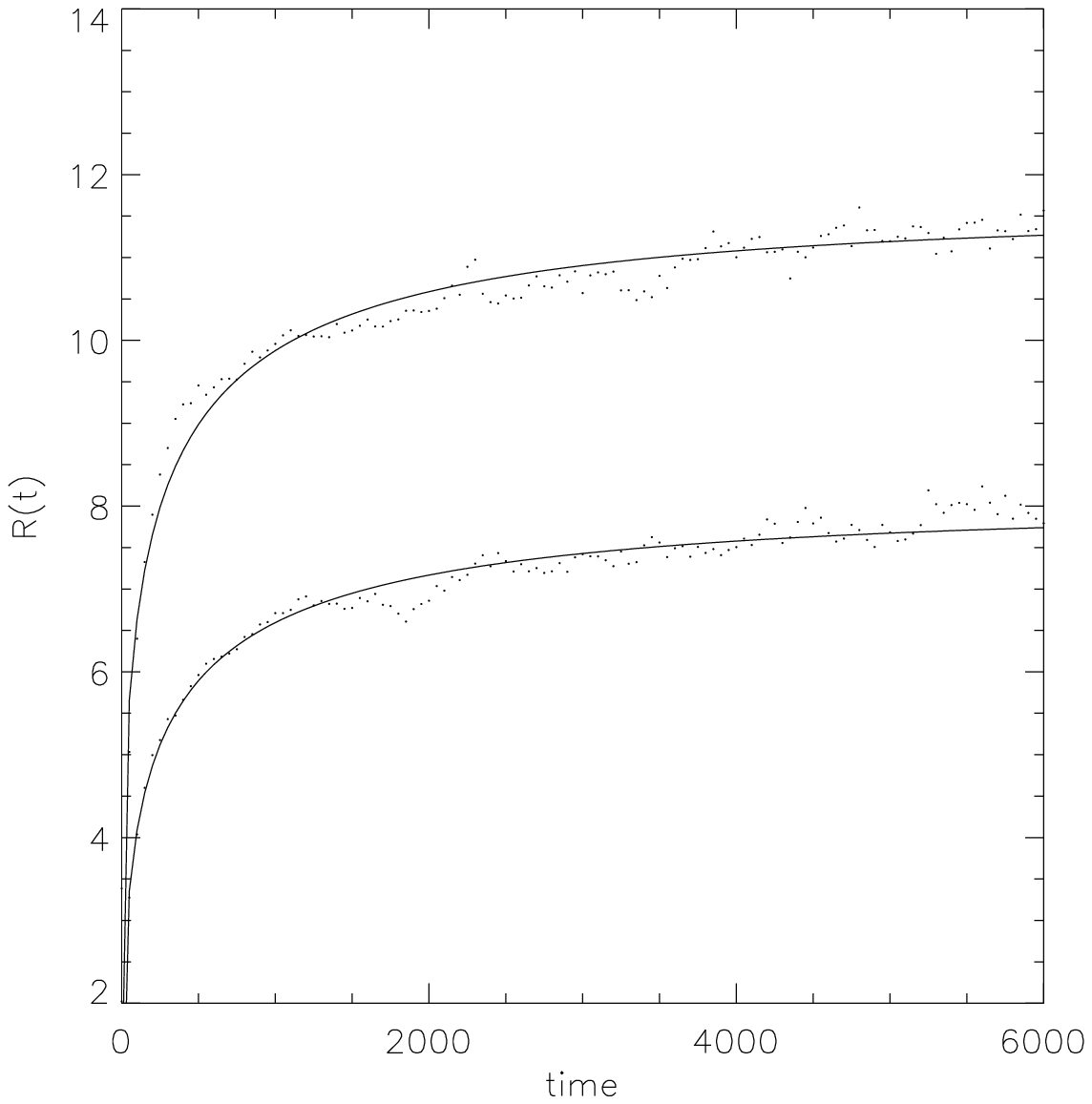}}
\end{center}
\caption{}
\label{fig:change}
\end{figure}

We investigated our results by visualization and plots of domain
size against timesteps on different scales. In all cases we could not
see algebraic growth even at early timesteps, indicating that the high
amount of surfactant used drastically changes the usual domain
growth. For the early-time behaviour we found logarithmic growth,
$R(t) \propto (\ln t)^\theta$, with growth exponents of $\theta = 0.9$
for $O = 0.04$, $\theta = 1.3$ for $O=0.08$ and $O=0.12$ and
$\theta = 1.45$ for $O=0.15$ and $O=0.18$. At late times, however,
the data points for all simulations but for $O = 0.08$, which will be
discussed below, are consistently below that function and move towards
a saturated value.
Fitting the data to the stretched exponential form gave good agreement
with the measured points, as shown in Fig.~\ref{fig:change}. This
indicates that the stretched exponential form is applicable in
describing the domain growth in a wide range of microemulsion phases.
For $O = 0.08$, however, we saw behaviour similar to our results in the
droplet phase with $\Gamma\approx 2.5$. Following an early-time
logarithmic growth, a nucleation process takes place after which we
observe very noisy behaviour. Visualization of these simulations also
revealed similarities, with the formation of two distinct
droplet-sizes and formation of water droplets in larger oil droplets. 
Visualization of the other simulations confirmed that at $O=0.04$ we
see stable circular oil droplets of one size. At $O = 0.12$ we find
that some of the droplets continue to grow and are deformed into
ellipses and stripes, but do not yet connect to form the larger structures
that are characteristic of the sponge phase. At $O=0.15$ the latter
structures have formed, and are difficult to distinguish from the case
$O=0.18$, where equal amounts of oil and water are present.

\section{Discussion and Conclusions}
\label{sec:dac}

We have studied both binary immiscible fluid and ternary amphiphilic
fluid dynamical behaviour in off-critical quenches using our
hydrodynamic lattice-gas model. In the binary case we found algebraic
scaling laws in the early and late time behaviour of the system over a
wide range of concentrations of the minority phase. In the late time
regime, our results give a growth exponent of $n = 0.45$ for
minority phase concentrations ranging from $20\%$ up to $40\%$. 
This is in agreement with other simulations
~\cite{bib:veto,bib:cn,bib:walc}, which all studied domain growth 
for one or two different concentrations, but is below the theoretical
prediction by Furukawa~\cite{bib:f} and San Miguel
{\em et al.}~\cite{bib:smgg}.
In the early time regime we do not find such a dominating scaling
regime. Instead we find a strong dependence of the growth exponent on the
concentration of the minority phase. Although noise interferes with our
results, we find a growth exponent of $n = 0.2$ as the lowest value in the
early time regime, slightly below the value of $n =\frac{1}{4}$,
predicted by a renormalization group approach~\cite{bib:ccz}. Results
from the ternary amphiphilic fluid case indicate that the lower
threshold is even below the value of $n = 0.2$, as the presence of
surfactant, which has shown to move the system towards early-time
growth~\cite{bib:ecb}, reduces the early-time growth exponent to
$n=0.14$. Possible reasons for the variation of $n$ are the existence
of two or more growth regimes or corrections of scaling regimes, that
lead to an asymptotic behaviour of the growth exponent.

In the ternary case with surfactant present, the kinetics of the
droplet phase self-assembly have been studied. Little is currently known
experimentally about amphiphilic kinetics, so simulations provide
the only way to deal with the effects that arise in these systems.
We find that the presence of amphiphilic molecules affects the
self-assembly kinetics very dramatically. For the early time
dynamics we find algebraic growth, the growth exponent diminishing as
the surfactant concentration is raised. At sufficiently high
surfactant concentrations, we find a second scaling
regime in which the surfactant hastens the process of domain
growth.  We were able to show that after this nucleation process two
length scales control the behaviour of our system. We found evidence
that the system is then in a coexistence region of the oil- and
water-droplet phases. This nucleation process has never been reported
before. At surfactant
concentrations above twice the oil concentration we find that
the growth law can be better described as logarithmic,
$R(t) \propto (\ln t)^\theta$. This behaviour has
been found in several earlier papers~\cite{bib:ecb,bib:lmtz}, which
studied the dynamics of the sponge phase, and might be a common
feature of emulsions. At surfactant concentrations that are
three times as high as the oil concentration, we find saturation of
the oil droplets that can be described well with a stretched
exponential function. As domain saturation occurs, the nucleation
mechanism is no longer seen in our simulations, indicating that the
stability of the droplet phase is now beyond the critical value at
which the system undergoes a transition into the coexistence region.

Finally we studied the transition between the droplet and the sponge
phases, where we demonstrated that our system consistently simulates
the correct behaviour in the transition between both of these limits.
The validity of the stretched exponential form during this
transition was shown. We were able to show that an accumulation of
relaxation times combined with an essential singularity in the
weighting function of the relaxation times can result in a stretched
exponential form.

Detailed experimental studies of amphiphilic self-assembly kinetics
would be valuable for comparison with the stretched exponential
behaviour found here, and to explore further the nature of the
nucleation process revealed in the present study.

\section*{Acknowledgements}
 
FWJW, PVC and BMB would like to thank Alan Bray, Colin Marsh, Andrew
Rutenberg and Julia Yeomans for stimulating discussions. FWJW is
indebted to the Bayerische Begabtenf\"{o}rderung and Balliol
College, Oxford University, for supporting his stay in Oxford.
BMB and PVC acknowledge NATO grant number CRG950356 and the CCP5
committee of EPSRC for funding a visit to U.K. by BMB. PVC is grateful
to Wolfson College and the Department of Theoretical Physics, Oxford
University, for a visiting fellowship (1996-1998).

\newpage

\section*{Figures}

\vspace{0.5cm} 

\noindent Figure 1: Temporal (time-steps) growth of domain size
(lattice units), $R(t)$, for $\beta = 0.5$, shown on a
logarithmic-scale plot. The concentrations of the minority phase 
are, from top to bottom: $44\% , 36\% , 24\% , 14\%$. The straight
lines are included as guides to the eye only and have slopes of
$0.55, 0.45, 0.45, 0.25$, respectively. In order to make the plots
more distinguishable, the ordinates have been multiplied by factors
$2.0, 1.5, 1.0, 0.5$, from top to bottom. 

\vspace{1cm}

\noindent Figure 2: Temporal (time-steps) growth of domain size
(lattice units), $R(t)$, for $\beta = 0.15$, shown on a
logarithmic-scale plot. The concentrations of the minority phase 
are, from top to bottom: $44\% , 39\%$. The straight
lines are included as guides to the eye only and have slopes of
$0.5$ and $0.33$, respectively. The upper ordinates have been
multiplied by a factor of $1.5$.

\vspace{1cm}

\noindent Figure 3: Temporal (time-steps) growth of domain size
(lattice units), $R(t)$, for binary fluid and $\beta = 0.137$, shown on
a logarithmic-scale plot. The concentrations of the minority phase 
are, from top to bottom: $44\% , 39\%$. The straight
lines are included as guides to the eye only and have slopes of
$0.45$ and $0.3$, respectively. The upper ordinates have been
multiplied by a factor of $1.5$.

\vspace{1cm}

\noindent Figure 4: Growth exponent $n$ plotted against the concentration
of the minority phase. Diamonds ($\Diamond$) denote the values for
$\beta = 0.5$, triangles ($\bigtriangleup$) are for $\beta = 0.15$ and
boxes ($\Box$) for $\beta=0.137$. The diagram is symmetric due to
the symmetry of the model with respect to the coloured particles.
Note that only the early-time exponents are plotted for the crossover
regime, where $\beta = 0.15$.

\vspace{1cm}

\noindent Figure 5: Temporal (time-steps) growth of domain size
(lattice units), $R(t)$, for a ternary amphiphilic fluid, shown on
a logarithmic-scale plot. The value of $\Gamma$ is 0.25
for the upper and 0.625 for the lower curve. The ordinates for the upper
curve have been multiplied by 1.5 in order to make the plots more
distinguishable. The straight lines are included as guides to the eye
only and have slopes of $0.45$ and $0.38$ respectively.

\vspace{1cm}

\noindent Figure 6: Temporal (time-steps) growth of domain size
(lattice units), $R(t)$, for a ternary amphiphilic fluid, shown on
a logarithmic-scale plot. The value of $\Gamma$ is, from
top to bottom, $1.125, 1.375$ and $2.0$. The ordinates for the upper
curve have been multiplied by 2, the ones for the lower curve by
$\frac{1}{2}$ in order to make the plots more distinguishable.
The straight lines are included as guides to the eye only and have slopes of
$0.28$ and $0.58$ for the upper, $0.25$ and $0.75$ for the middle and
$0.14$ and $0.90$ for the lower curve.

\vspace{1cm}

\noindent Figure 7: Temporal evolution of a ternary system with
$\Gamma = 1.5$. Oil majority sites are red, water sites
are black and surfactant sites are green. Note that the nucleation sets in
between 3000 and 4000 timesteps. 

\vspace{1cm}

\noindent Figure 8: Temporal evolution of different droplet sizes 
(lattice units), $R(t)$, for a ternary amphiphilic fluid, shown on
a logarithmic-scale plot. The value of $\Gamma$ is $1.125$. Boxes ($\Box$)
show all droplets with radii between $2$ and $50$, triangles ($\bigtriangleup$)
are for droplets with radii between $5.5$ and $50$ and
diamonds ($\Diamond$) mark droplets with radii between $2$ and $11$.

\vspace{1cm}

\noindent Figure 9: Temporal (time-steps) growth of domain size
(lattice units), $R(t)$, for a ternary amphiphilic fluid, shown on
a logarithmic-scale plot. The value of $\Gamma$ is  $ 2.375$
The straight line is included as guide to the eye only and has a slope of
$0.15$

\vspace{1cm}

\noindent Figure 10: Growth of domain size (lattice units), $R(t)$, for
a  ternary amphiphilic fluid against $\ln t$ (time-steps), shown on a
logarithmic-scale plot. 
The value of $\Gamma$ is, from top to bottom, $2.375$ and $2.875$. 
The ordinates of the upper curve have been multiplied by 1.5 in order
to make the plots more distinguishable.
The straight lines are included as guides to the eye only and have slopes of
$1.0$ and $0.8$ respectively.

\vspace{1cm}

\noindent Figure 11: Growth of domain size against $\ln t$, shown on
a logarithmic-scale plot. The value of $\Gamma$ is $3.0$.
The straight line is included as guide to the eye only and has
a slope of $0.75$

\vspace{1cm}

\noindent Figure 12: Growth of domain size (lattice units) against $t$
(time-steps), shown on a linear-scale plot. 
The results are from the study of the droplet phase; the value
of $\Gamma$ is $3.25$ for the upper and $3.75$ for the lower curve.
The ordinates of the upper curve have been multiplied by 1.5. The
solid lines are stretched exponential fits to the data.

\vspace{1cm}

\noindent Figure 13: Growth of domain size (lattice units) against $t$
(time-steps), shown on a linear-scale plot. 
The results are from the study of the transition between sponge and
droplet phase; the value of the reduced density of oil is
$O=0.15$ for the upper and $O=0.12$ for the lower curve.
The ordinates of the upper curve have been multiplied by 1.5. The
solid lines are stretched exponential fits to the data.

\newpage

\section*{Tables}

\vspace{0.1cm} 

\noindent Table I: Results from the nucleation regime. $T_{nuc}$ is the
timestep at which nucleation occurs, $R_{nuc}$ the average domain
size, $n_e$ and $n_l$ are the algebraic growth exponents in the early
and late time regime, respectively, and $\Gamma$ is the
surfactant:oil ratio.

\vspace{1.0cm}

\begin{table} 
\begin{center}
\begin{tabular}{|c|c|c|c|c|} \hline
\em{$\Gamma$} & \em{$T_{nuc}$} & \em{$R_{nuc}$} & \em{$n_e$} &
\em{$n_l$} \\ \hline
$1.0$    &  $2000$ & $8$ & $0.3$ & $0.5$ \\
$1.125$  &  $1700$ & $8$ & $0.28$ & $0.58$ \\
$1.25$   &  $1800$ & $7$ & $0.25$ & $0.58$ \\
$1.375$  &  $3500$ & $8$ & $0.25$ & $0.75$ \\
$1.5$    &  $3500$ & $7.5$ & $0.22$ & $0.78$ \\
$1.75$   &  $3000$ & $6$ & $0.19$ & $0.8$ \\
$2.0$    &  $4000$ & $5$ & $0.14$ & $0.9$ \\  \hline
\end{tabular}
\end{center}
\caption{}
\label{tab:nuc}
\end{table}

\end{document}